%% file: main.tex
\documentclass[twocolumn,final,twoside,journal]{IEEEtran}

\input{./tex_files/packages}

\input{./tex_files/notation}

\usepackage{multirow}

\begin{document}
\input{./tex_files/acron}

\input{./tex_files/title}
\input{./tex_files/intro}

\input{./tex_files/prelim}
\input{./tex_files/algs}

\input{./tex_files/DE}

%\input{./tex_files/BSC}

\input{./tex_files/results}

\input{./tex_files/conclusions}

%\begin{appendices}
\appendix
\input{./tex_files/appendix}

\balance
\bibliographystyle{IEEEtran}
\bibliography{IEEEabrv,references}

%\input{./tex_files/reworked_notation}

%\atColsEnd{\vskip-1pt}
%\hspace{1cm}

\end{document}

%% file: tex_files/packages.tex
\usepackage{amsmath,amsthm,amssymb,graphicx,cite,bm}
\usepackage{epsfig,amsfonts}
\usepackage{graphicx}
\usepackage{color}
\usepackage[dvipsnames]{xcolor}
\usepackage[nolist]{acronym}
\usepackage{psfrag}
\usepackage{perso}

\usepackage{pgfplots}
 \pgfplotsset{compat=newest}
    \pgfplotsset{plot coordinates/math parser=false}
    \pgfplotsset{
    label style={anchor=near ticklabel},
    xlabel style={yshift=0.0em},
    ylabel style={yshift=-0.3em},
    tick label style={font=\footnotesize },
    label style={font=\footnotesize},
    legend style={font=\footnotesize},
    title style={font=\fontsize{7}}}

\usepackage{mathtools}
\mathtoolsset{showonlyrefs}

\usepackage{placeins}

\usepackage{multirow}
\usepackage{etoolbox}
\usepackage{hhline}

\usepackage{ulem,soul}
\usepackage{balance} 
\usepackage{ctable}

%% file: tex_files/notation.tex
\newtheorem{remark}{Remark}

\newcommand{\dv}{\mathsf{d_{v}}}
\newcommand{\dc}{\mathsf{d_{c}}}

\newcommand{\inputalphabet}{\mathcal{X}}
\newcommand{\inputrv}{X}
\newcommand{\inputrvreal}{x}
\newcommand{\outputalphabet}{\mathcal{Y}}
\newcommand{\outputrv}{Y}
\newcommand{\outputrvreal}{y}

\newcommand{\proberror}{\epsilon}

\newcommand{\Rc}{R}

\newcommand{\ensC}[3]{\mathscr C_{ #1, #2}^{#3}}

\newcommand{\vn}{\mathtt{v}}%variable node

\newcommand{\cn}{\mathtt{c}}%check node

\newcommand{\vcm}[2]{p_{#1}^{(#2)}}%pr  of var to check node message
\newcommand{\cvm}[2]{s_{#1}^{(#2)}}
\newcommand{\neigh}[1]{\mathcal{N} ( #1) }

\newcommand{\mes}[3]{m_{ #1 \rightarrow #2 }^{(#3)}}
\newcommand{\Mes}[3]{M_{ #1 \rightarrow #2 }^{(#3)}}

\newcommand{\mesch}[1]{m_{ #1 }}

\newcommand{\Fq}{\mathbb{F}_q}

\renewcommand{\argmax}[1]{\underset{#1}{\mathrm{arg \, max}}\,}

\ifdef{\l}
{\renewcommand{\l}[1]{L_{ #1}} }
{\newcommand{\l}[1]{L_{ #1}}}

\newcommand{\genfq}{b}

\newcommand{\F}[1]{F_{#1}}
\newcommand{\f}[1]{f_{#1}}

\newcommand{\Fv}{\bm{F}}
\newcommand{\fv}{\bm{f}}

\newcommand{\I}{\mathbb{I}}

\renewcommand{\a}{a}
\renewcommand{\b}{y}

\newcommand{\agg}{E}
\newcommand{\aggc}[2]{\agg_{#1}^{(#2)}}
\newcommand{\aggv}[1] {\bm{\agg}^{(#1)}}
\newcommand{\lmaxim}[1]{\mathcal{\agg}^{(#1)}}
\newcommand{\aggapp}[1]{L^{\mathsf{APP}}_{#1}}
\newcommand{\aggappv}{\bm{L}^{\mathsf{APP}}}

\ifdef{\H}
{\renewcommand{\H}{\mathsf{U}}}
{\newcommand{\H}{\mathsf{U}}}

\ifdef{\D}
{\renewcommand{\D}{\mathsf{D}}}
{\newcommand{\D}{\mathsf{D}}}

\newcommand{\nel}[1]{{\mu}_{#1} } %
\newcommand{\nmaxr}{\kappa}
\newcommand{\nmaxrmax}{\kappa_{\mathrm{max}}}

\newcommand{\accvote}[1]{{t_{#1}}}
	
\newcommand{\az}{\accvote{0}}
\newcommand{\ao}{\accvote{1}}

%% file: tex_files/acron.tex
\begin{acronym}
\acro{BSC}{binary symmetric channel}
\acro{BMP}{binary message passing}
\acro{BP}{belief propagation}
\acro{DE}{density evolution}
\acro{LDPC}{low-density parity-check}
\acro{MDPC}{moderate-density parity-check}
\acro{lv}{$L$-value}
\acro{llv}{$\bm L$-vector}
\acro{LLR}{low-likelihood ratio}
\acro{QSC}[$q$-SC]{$q$-ary symmetric channel}
\acro{SER}{symbol error rate}
\acro{SMP}{symbol message passing}
\acro{VN}{variable node}
\acro{CN}{check node}
\acro{RV}{random variable}
\end{acronym}

%% file: tex_files/title.tex
\title{Symbol Message Passing Decoding of Nonbinary Low-Density Parity-Check Codes}

\author{
    \IEEEauthorblockN{Francisco L\'azaro\IEEEauthorrefmark{2}, Alexandre Graell i Amat$^\ddag$, Gianluigi Liva\IEEEauthorrefmark{2}, Bal\'azs Matuz\IEEEauthorrefmark{2}\\
    \IEEEauthorblockA{}\IEEEauthorrefmark{2}Institute of Communications and Navigation of DLR (German Aerospace Center),
    Wessling, Germany.}\\% Email: \{Francisco.LazaroBlasco, Gianluigi.Liva, Balazs.Matuz\}@dlr.de}\\
	$^\ddag$Department of Electrical Engineering, Chalmers University of Technology,
    Gothenburg, Sweden.
    \vspace{-1cm}
    
\thanks{ This work will be presented at IEEE Globecom 2019
	
	\copyright 2019 IEEE. Personal use of this material is permitted. Permission
	from IEEE must be obtained for all other uses, in any current or future media, including
	reprinting /republishing this material for advertising or promotional purposes, creating new
	collective works, for resale or redistribution to servers or lists, or reuse of any copyrighted
	component of this work in other works}} 
%}
\maketitle

\thispagestyle{empty} \pagestyle{empty}

\begin{abstract}
We present a novel decoding algorithm  for $q$-ary \acl{LDPC} codes, termed \acl{SMP}.
The proposed algorithm can be seen as a generalization of Gallager B and the \acl{BMP}  algorithm by Lechner \emph{et al.} to $q$-ary codes.
We derive density evolution equations for the \acl{QSC}, compute thresholds for a number of regular \acl{LDPC} code ensembles, and verify those by Monte Carlo simulations of long channel codes. 
The proposed algorithm shows performance advantages with respect to an algorithm of comparable complexity from the literature.
\end{abstract}

\vspace{-0.25cm}

%%%%%%%%%%%%%%%%%%%%%%%%%%%%%%%%%%%%%%%%%%%%%%%%%%%%%%%%%%%%%%%%%%%%%%%%%%%%%%%%%%%%%%%%%%%%%%%%%%%%%%%%%%

%\begin{keywords}
%\end{keywords}

%%%%%%%%%%%%%%%%%%%%%%%%%%%%%%%%%%%%%%%%%%%%%%%%%%%%%%%%%%%%%%%%%%%%%%%%%%%%%%%%%%%%%%%%%%%%%%%%%%%%%%%%%% 

%% file: tex_files/intro.tex
\section{Introduction}\label{sec:Intro}

%Low-density parity-check codes \acused{LDPC} (\ac{LDPC}) were first introduced by Gallager in his PhD thesis \cite{Gallager63}.
%Despite being largely forgotten for a long period of time, they are nowadays used in various communication standards \cite{DVB-S2:ch,5G:ch}. 
%Recently, binary \ac{LDPC}-like codes have been proposed for their use in McEliece-like cryptosystems \cite{mceliece1978public}, \cite{mdpc,baldi2018ledakem}. The codes used are sometimes of moderate rather than low density, leading to the term \ac{MDPC} codes. In this context, low decoding complexity is of tantamount importance, and hence so called bit-flipping decoding algorithms are used, in which the messages exchanged between check and variable nodes take values in a binary alphabet.
There is a large body of literature considering message passing algorithms for binary \ac{LDPC} codes. In his seminal work \cite{Gallager63}, Gallager proposed two different message passing algorithms for LDPC codes, nowadays known as Gallager A and B, which exchange \emph{binary} messages between \acp{CN} and \acp{VN}.  In \cite{richardson2001capacity}, algorithm E was proposed, where messages take values in a ternary alphabet. %A particularly interesting algorithm known as \ac{BMP} was introduced in \cite{Lechner:BMP}. 
A powerful algorithm, referred to as \ac{BMP} was introduced in \cite{Lechner:BMP}. Although the exchanged messages are binary, the algorithm is able to exploit soft information from the channel at the \acp{VN}. An extension of \ac{BMP} to ternary message alphabets was studied in \cite{yacoub2018protograph}.  A finite alphabet message iterative decoder for the \ac{BSC} was presented in \cite{planjery2013finite}.

%Low-complexity decoding algorithms for $q$-ary \ac{LDPC} codes have received considerably less attention than their binary counterparts, despite the fact that a message passing algorithm was already proposed by Gallager in  \cite{Gallager63}.  

Various works in the literature study the extension of binary \ac{LDPC} codes to larger fields, including the original work by Gallager \cite{Gallager63}. Nonbinary \ac{LDPC} codes constructed over finite fields for binary-input Gaussian channels were investigated in \cite{Mac_Kay_LDPC_GFq_1998}. Different simplified message passing algorithms were studied in \cite{Declerq:FFT_dec_2003,declercq2007decoding}. Regarding \acp{QSC},
a majority-logic-like decoding algorithm was introduced in \cite{metzner1996majority}, while verification based decoding algorithms were studied in \cite{luby2005verification,MLP+13,shokrollahi2004low,zhang2011analysis}. Both algorithms target large field orders. 
In \cite{kurkoski2007density} a list message passing decoding algorithm for $q$-ary \ac{LDPC} codes over the \ac{QSC} was proposed, which is practical when the list size is small. For list size 1, the exchanged messages take values in a $(q+1)$-ary message alphabet, composed of the elements of $\Fq$ and an additional erasure message.
In \cite{stark2018decoding} a decoding algorithm for $q$-ary \ac{LDPC} codes was presented, for which the \ac{CN} and \ac{VN} operations are implemented by means of look up tables. It makes use of the information bottleneck method and is practical for small $q$.

This paper targets $q$-ary \ac{LDPC} codes for which we propose a low-complexity decoding algorithm,  termed \ac{SMP}. The proposed algorithm can be seen as an extension of 
\ac{BMP} to $q$-ary codes and $q$-ary message alphabets. Similarly to \ac{BMP}, it can exploit soft information from the channel at the \acp{VN}. Over the \ac{QSC}, \ac{SMP} becomes a natural 
generalization of Gallager B  \cite{Gallager63}. 
We develop a \ac{DE} analysis for \ac{SMP} over the \ac{QSC}.
% and, as a byproduct, yield the reliability of the extrinsic channels, as it will be explained in Section~\ref{sec:DE}}.
For large $q$, the evaluation of \ac{DE} becomes infeasible, due to the increasing complexity. To tackle this, we derive tight upper and lower bounds on the iterative decoding thresholds, which can be efficiently evaluated even for very large $q$. 
Simulation results are compared with the decoding thresholds obtained via \ac{DE}. Both the analysis and the simulations are provided for the case of regular \ac{LDPC} code ensembles for ease of exposition. However, the extension to irregular ensembles is straightforward. For the considered ensembles, the derived thresholds are superior to the ones obtained in \cite{kurkoski2007density} with list size $1$. 

The proposed algorithm is of interest, among others, for applications with high decoding throughput and low decoding complexity requirements, such as  optical communications. Another application area is code-based post-quantum cryptography, for which binary regular \ac{LDPC} codes are considered in the literature \cite{mdpc}. Nonbinary codes can render cryptanalysis more difficult, but there is the need for simple decoders.

%	where codes over large fields can render cryptanalysis more complex, and low complexity decoding algorithms are required.} \fran{In this work, we focus on unstructured regular \ac{LDPC} ensembles, since this is the setting considered . However, an extension irregular ensembles is straightforward. }

%The rest of the paper is organized as follows. Section~\ref{sec:Prelim} presents the notation and introduces and some preliminary definitions. In Section~\ref{sec:Algs} a formal definition of \ac{SMP} over the \ac{QSC} is presented. Section~\ref{sec:DE} introduces the density evolution analysis over the \ac{QSC}. Numerical results are presented in Section~\ref{sec:results} and conclusions follow in Section~\ref{sec:conclusions}. 

%% file: tex_files/prelim.tex
\section{Preliminaries}\label{sec:Prelim}

In this work, we consider regular $(\dv, \dc )$ \ac{LDPC} codes constructed over a finite field of order $q$, $\Fq$. The code's bipartite graph comprises $n$ \acp{VN} $\vn_j$, $j=\{1,2,\hdots, n\}$ 
of degree $\dv$ and $m$ \acp{CN} $\cn_i$, $i=\{1,2,\hdots, m\}$ of degree $\dc$. The design rate is $\Rc=1-m/n=1-\dv/\dc$. 
%An edge in the bipartite graph connecting  $\vn$ to  $\cn$  possesses an edge label $h_{\vn,\cn} \in \Fq \setminus 0$.  
The edge label associated to the edge connecting  $\vn$ and  $\cn$ is denoted by $h_{\vn,\cn}$, with $h_{\vn,\cn} \in \Fq \setminus 0$.
The neighborhood of a \ac{VN}, i.e., the set of all connected \acp{CN}, is denoted as $\neigh{\vn}$.
%\footnote{The connections between \acp{VN} and \acp{CN} and the edge labels are determined by the code's parity-check matrix (see, e.g., \cite{declercq2007decoding}).} 
Similarly, the neighborhood of a \ac{CN} is denoted as $\neigh{\cn}$. At the 
$\ell$th decoding iteration, let the message sent from  $\vn$ to  $\cn$ be $\mes{\vn}{\cn}{\ell}$, and the message from  $\cn$ to  $\vn$ be $\mes{\cn}{\vn}{\ell}$. Furthermore, the channel observation at $\vn$ is denoted by $\mesch{\vn}$. The ensemble of $q$-ary regular $(\dv, \dc )$ codes with block-length $n$ is denoted by  $\ensC{\dv}{\dc}{q}$ and is defined by a uniform distribution over all possible edge permutations between \acp{VN} and \acp{CN} and over all possible edge labelings from $\Fq \setminus 0$.

Consider a \ac{QSC} with error probability $\proberror$, input alphabet $\inputalphabet$ and output alphabet $\outputalphabet$, with  ${\inputalphabet = \outputalphabet = \{0, \alpha^0, \hdots, \alpha^{q-2}\}}$, where $\alpha$ is a primitive element of $\Fq$. 
Denote by $\inputrv \in \inputalphabet$  and $\outputrv \in \outputalphabet$ the \acp{RV} associated to the channel input and channel output, respectively, and by $\inputrvreal$ and $\outputrvreal$ their realizations. 
Then, the transition probabilities of the \ac{QSC} are
\begin{equation}
P_{Y|X}(y|x) =  \begin{cases}
                                                            1 - \proberror & \mbox{if } \outputrvreal = \inputrvreal \\
                                                            \proberror / (q-1) & \mbox{otherwise}.
                                                          \end{cases} \label{eq:QSC_trans_prob}
\end{equation}
The capacity of the \ac{QSC}, in symbols per channel use, is
\begin{equation}\label{eq:capacity_q}
C= 1 + \proberror \log_q \frac{\proberror}{q-1} + (1-\proberror) \log_q (1-\proberror).
\end{equation}

For a given channel output $y$, we introduce the normalized log-likelihood vector, also referred to as \acl{llv},
\begin{equation}\label{eq:llvector}
\bm{L}(y)=\left[ L_0(y), L_1(y), \ldots , L_{\alpha^{q-2}}(y)\right]
\end{equation}
whose elements are obtained as
\[
L_\genfq(y)=\log \left(P_{Y|X}(y|b) \right) - \log\left( \proberror / (q-1) \right) .
\]
From \eqref{eq:QSC_trans_prob}, we have
\begin{equation} \label{eq:Li_QSC}
L_\genfq(y)= \begin{cases}
\D(\proberror)& \mbox{if } \genfq = y \\
0 & \mbox{otherwise}
\end{cases}
\end{equation}
where 
\[
\D(\proberror)= \log(1-\proberror) - \log\left( \proberror / (q-1) \right).
\]

%In the next two sections we consider a \ac{QSC} as communication channel model. In Section~\ref{sec:BSC} we change the channel model to a \ac{BSC}, construct a $q$-ary super channel, and illustrate how \ac{SMP} can be adapted for such channels.

%Given a vector $\mathbf{\vecgen}=(\vecgen_1, \vecgen_2, \dots, \vecgen_h) \in \Fq^h$, we define its composition $\comp(\mathbf{\vecgen})$ as
%\[
%\comp(\mathbf{\vecgen}) = \left[ \comp_0(\mathbf{\vecgen}), \comp_1(\mathbf{\vecgen}), \dots,  \comp_{q-1}(\mathbf{\vecgen})  \right]
%\]
%where
%\[
%\comp_i(\mathbf{\vecgen}) = \left| \left\{ \vecgen_j:\vecgen_j = \alpha^{i-1} \right\} \right|, \,\, \text{for } i \in \{1, 2, \dots, q-1\}
%\]
%being $\alpha$ the residue class of the polynomial $x$, and
%\[
%\comp_0(\mathbf{\vecgen}) = \left| \left\{ \vecgen_j:\vecgen_j =0 \right\} \right|.
%\]
%Thus, the composition of a vector simply tells us how many of its elements are equal to each of the elements in $\Fq$. 

%% file: tex_files/algs.tex
\section{Symbol Message Passing Decoding}\label{sec:Algs}

%\subsection{Algorithm Description}
In this section, we describe the proposed \ac{SMP} algorithm in detail, assuming transmission over the \ac{QSC}. \ac{SMP} decoding is an iterative algorithm, where \acp{CN} and \acp{VN} exchange  $q$-ary messages. The basic steps of \ac{SMP} are as follows.

\begin{itemize}
	\item[i.] \textbf{Initialization.}
	At the first iteration, each \ac{VN} $\vn$ sends to all $\cn \in \neigh{\vn}$
	\[
	\mes{\vn}{\cn}{1}=\mesch{\vn}
	\]
	where $\mesch{\vn}=y$, $y$ being the channel observation associated to \ac{VN} $\vn$.
	\item[ii.] \textbf{\ac{CN}-to-\ac{VN} step.} Each \ac{CN} computes
	\begin{equation}
	\label{eq:check_node_GA}
	\mes{\cn}{\vn}{\ell} = h_{\vn,\cn}^{-1} \sum_{ \vn' \in \neigh{\cn} \setminus \vn} h_{\vn',\cn} \, \mes{\vn'}{\cn}{\ell} .
	\end{equation}
	%where $h_{\vn,\cn}^{-1}$ is the multiplicative inverse of $(\vn,\cn)$ edge label. %, i.e., the parity-check matrix element corresponding to the $(\vn,\cn)$ edge in the corresponding bipartite graph.
	\item[iii.]  \textbf{\ac{VN}-to-\ac{CN} step.} Let $\aggv{\ell}$ be an aggregated extrinsic \acl{llv}, with
	\begin{align} \label{eq:E-vector} 
	\aggv{\ell}&=\left[\aggc{0}{\ell},\aggc{1}{\ell},\ldots,\aggc{\alpha^{q-2}}{\ell} \right]\\ \label{eq:LL-vector}
	&= \bm{L}\left(\mesch{\vn}\right) + \sum_{\cn' \in \neigh{\vn} \setminus \cn}  \bm{L}\left(\mes{\cn'}{\vn}{\ell-1}\right) .
	\end{align}	
	%
	%where the  \aclp{llv} are obtained from $\eqref{eq:Li_QSC}$. 
	Then, each \ac{VN} computes
	\[
	\mes{\vn}{\cn}{\ell} = \argmax{\genfq \in \Fq}  \aggc{\genfq}{\ell} .
	\]
	Whenever multiple maximizing arguments exist, the $\arg\,\max$ function returns one of them at random with uniform probability. The \ac{VN} operation can be interpreted as if the \acp{CN} and the channel would \emph{vote} for the value of the code symbol associated to the \ac{VN}.
	The \ac{VN} assigns different weights to the \ac{CN} and channel votes and selects the element  with the highest score.

	In \eqref{eq:LL-vector}, the \acl{llv}  corresponding to the channel observation is obtained from $\eqref{eq:Li_QSC}$ using the channel error probability $\proberror$. 
	Further, we model the \ac{CN}-to-\ac{VN} messages, as an observation of the symbol $X$ (associated to $\vn$), at the output of an \emph{extrinsic} \ac{QSC} channel \cite{ashikhmin2004extrinsic,Lechner:BMP}. The extrinsic channel error probability is denoted by $\xi^{(\ell)}$ and is used to compute the corresponding \aclp{llv} in \eqref{eq:LL-vector}. 
 In general, the error probabilities $\xi^{(\ell)}$ are not known. Estimates can be obtained from \ac{DE} analysis, as proposed in \cite{Lechner:BMP,yacoub2018protograph}.
	
	\item[iv.] \textbf{Final decision.} 
	After iterating steps ii.\ and iii.\ for $\ell_{\max}$ iterations, the final decision at each \ac{VN} is computed as
	\begin{align}
	\hat{x} = \argmax{\genfq \in \Fq} ~ \aggapp{\genfq}
	\end{align}
	with
	\begin{align}
		\aggappv&=\left[\aggapp{0},\aggapp{1},\ldots,\aggapp{\alpha^{q-2}}\right]\\ \label{eq:app}
		&= \bm{L}\left(\mesch{\vn}\right) + \sum_{\cn \in \neigh{\vn}}  \bm{L}\left(\mes{\cn}{\vn}{\ell_{\max}}\right).
	\end{align}
\end{itemize}

\subsection{Complexity Analysis}
%\fran{The complexity of iterative decoding algorithms depends on two aspects: (i) the complexity of the \ac{VN} and  \ac{CN} update rules, and (ii) the data flow between \acp{CN} and \acp{VN}, which depends on the size of the messages exchanged. 
%}

The complexity of \ac{SMP} is implementation dependent and can be studied from many perspectives. Here, we focus on the data flow in the decoder, as well as on the number of arithmetic operations per iteration.

The internal decoder data flow, defined as the number of bits that are passed in each  iteration between \acp{VN} and \acp{CN}, is given by
$2\cdot n \cdot B \cdot  \dv$,
where $B$ is the number of bits used to represent each message. \ac{SMP} is characterized by a reduced data flow  between \acp{CN} and \acp{VN} compared to the classical \ac{BP} decoders for nonbinary \ac{LDPC} codes \cite{Mac_Kay_LDPC_GFq_1998, Declerq:FFT_dec_2003}. In \ac{SMP} all decoder messages are symbols in $\Fq$, rather than  $(q-1)$-ary probability vectors. It follows that $B=\log_2 q$ for \ac{SMP}, while for conventional nonbinary \ac{BP} decoding $B$ equals $(q-1)$ times the number of bits used to represent each probability. 

\ctable[
caption =SMP operations per iteration.,
label   =table:complexity,
pos     = t,
doinside=\footnotesize
]
{lccc}{}
{ 	\toprule
	Operation &  \ac{CN}   & \ac{VN} \\ 
	\midrule
	Addition, $\Fq$  &  $2\dc-1$   & -   \\
	Addition, real &  -  & $2\dv$   \\
	Multiplication, $\Fq$&  $2\dc$   & -   \\
	Maximization, real &  -   & $\dv$ \\ 
	\bottomrule
}

The algorithmic complexity of \ac{SMP} is summarized in Table~\ref{table:complexity} and is derived as follows. Consider the \ac{CN} update in \eqref{eq:check_node_GA}. Each incoming and outgoing message is multiplied by an element in  $\Fq \setminus 0$, yielding in total $2 \dc$ multiplications per \ac{CN}. One may precompute the sum of \emph{all} $\dc$ incoming messages, $h_{\vn',\cn} \, \mes{\vn'}{\cn}{\ell}$, $\vn' \in \neigh{\cn}$. Then, the extrinsic  message in \eqref{eq:check_node_GA} for an edge $(\cn,\vn)$ is obtained by subtracting the incoming message on that edge from the sum. This yields in total $\dc-1+\dc=2\dc -1 $ additions/subtractions, which are assumed to have equivalent in cost. %The same trick can be applied also at the \acp{VN}. 

At the \ac{VN} side, one may compute the sum of all $\dv+1$ \aclp{llv} in \eqref{eq:LL-vector} with only $\dv$ additions. {Note from \eqref{eq:Li_QSC} that a $q$-ary \acl{llv}  contains only a single non-zero element.} To obtain any of the $\dv$ extrinsic messages $\bm{\agg}$, the respective incoming \acl{llv} is subtracted from the sum. It follows that at each \ac{VN} the evaluation of \eqref{eq:LL-vector} can be implemented with  $\dv+\dv=2\dv$ additions/subtractions. Finally, for each of the $\dv$ extrinsic messages a maximum has to be found. %{Note that also $\bm{\agg}$ permits a compact representation with at most $\min\left(q,\dv\right)$ non-zero elements.}  %In fact, $\bm{\agg}$ contains $\dv-1$ (scaled) votes from the \acp{CN} and one (scaled) vote from the communication channel which might be the same or different.
The complexity of the proposed algorithm is very similar to the one of the algorithm in \cite{kurkoski2007density}, when the latter is operated with list size $1$.

%% file: tex_files/DE.tex
\section{Density Evolution Analysis}\label{sec:DE}
In this section we derive a \ac{DE} analysis for regular unstructured \ac{LDPC} code ensembles.
Due to the channel symmetry, without loss of generality, we assume that the all-zero codeword is transmitted. %\footnote{This is a common assumption if the error probability does not depend on the transmitted codeword\cite{richardson2001capacity}.}
We are interested in the probability  that the  \ac{RV} $\Mes{\vn}{\cn}{\ell}$ associated to the \ac{VN}-to-\ac{CN} message takes value $\a$ at the $\ell$th iteration, conditioned to the corresponding codeword symbol being zero,
	%\[
	%\vcmlong{\a}{\ell}.
	%\] 
%For the sake of notational simplicity, we introduce the shorthand notation
\[
	\vcm{\a}{\ell} =	\Pr\left\{	\Mes{\vn}{\cn}{\ell}=\a \big| \inputrv=0\right\}.
\] 
The initial probabilities $\vcm{\a}{0}$ are 
\[
\vcm{0}{0} = 1-\proberror
\]
and
\[
\vcm{\a}{0} = \proberror /(q-1), \qquad \forall \a \in \Fq \setminus 0.
\]
The iterative decoding threshold of a code ensemble  $\ensC{\dv}{\dc}{q}$ is defined as the maximum channel parameter $\proberror^{\star}$, so that for all $\proberror < \proberror^{\star}$,  $\vcm{0}{\ell}$ tends to $1$ as the block-length $n$ and the number of iterations $\ell$ tend to infinity \cite{richardson2001capacity}.

\begin{remark}As for the message passing algorithms proposed in \cite{Lechner:BMP,yacoub2018protograph}, \ac{DE} analysis plays a two-fold role. On one hand, it allows deriving the iterative decoding threshold of the \ac{LDPC} code ensemble under analysis. On the other hand, the analysis provides as a byproduct through \eqref{eq:extrinsic_reliability} estimates of the extrinsic channel reliabilities $\xi^{(\ell)}$ to be used in  step iii. of the decoding algorithm. The estimates turn to be accurate when decoding is applied to long codes (this is in fact the regime in which \ac{DE} analysis captures well the evolution of the message probability distributions).\end{remark}

Let $\cvm{\a}{\ell}$ be the probability that a \ac{CN}-to-\ac{VN} message takes value $\a$ at the $\ell$th iteration. We have
%Let us now consider the check to variable node messages. We have
\begin{equation}\label{eq_c_v}
\cvm{\a}{\ell} = \sum_{j=0}^{\dc-1} \binom{\dc-1}{j} \left(1-\vcm{0}{\ell}\right)^j \left(\vcm{0}{\ell}\right)^{\dc-1-j}  \psi_{j,\a}
\end{equation}
where $\psi_{j,\a}$ is the probability that $j$ erroneous messages sum up to $a$. Under the all-zero codeword assumption, the  extrinsic channel at the \ac{VN} input is a \ac{QSC} with error probability 
\begin{equation}
 \xi^{(\ell)} =1-\cvm{0}{\ell} \label{eq:extrinsic_reliability} .
\end{equation}  
The probability that $j$ independent \acp{RV} defined over $\Fq$, with zero probability assigned to the $0$ symbol and with uniform probability mass function over $\Fq \setminus 0$, sum up to zero is \cite[Appendix A]{lazaro2018bounds}
\[
\psi_{j,0} = \frac{1}{q} \left( 1 + \frac{(-1)^j}{(q-1)^{j-1}}\right).
\]
Due to symmetry, for any $\a \neq 0$, we obtain
\begin{align}\label{eq:check2var}
\psi_{j,\a}= \frac{1-\psi_{j,0}} {q-1} = \frac{1}{q} \left( 1 - \frac{(-1)^j}{(q-1)^{j}}\right).
\end{align}

Let us consider next the \ac{VN}-to-\ac{CN} messages. Define the random vector 
$\Fv^{(\ell)}$, 
\[
\Fv^{(\ell)}=\left(  \F{0}^{(\ell)},  \F{1}^{(\ell)} , \dots, \F{\alpha^{q-2}}^{(\ell)} \right)
\]
and its realization  $\fv^{(\ell)}$,
\[
\fv^{(\ell)}=\left(  \f{0}^{(\ell)},  \f{1}^{(\ell)} , \dots, \f{\alpha^{q-2}}^{(\ell)} \right)
\]
where $\F{\a}^{(\ell)}$ denotes the \ac{RV} associated to the number of \ac{CN}-to-\ac{VN} messages that take value $\a$ at the $\ell$th iteration, and $\f{\a}^{(\ell)}$  is its realization. The elements $\aggc{b}{\ell}$ of the aggregated extrinsic \acl{llv}  in \eqref{eq:E-vector} are related to $\f{b}^{(\ell)}$ and the channel observation $y$ by
\begin{equation}
\agg^{(\ell)}_\genfq = \begin{cases}
\D(\xi^{(\ell-1)}) \f{\genfq}^{(\ell-1)} + \D(\proberror) & \mbox{if } \genfq = y \\
\D(\xi^{(\ell-1)}) \f{\genfq}^{(\ell-1)}  & \mbox{otherwise.}
\end{cases}
\end{equation} 
Further, $\Fv^{(\ell)}$ conditioned to $\inputrv=0 $ is multinomially  distributed, with
\begin{align}\label{eq:multinom}
P_{\Fv^{(\ell)} | \inputrv } \left( \fv^{(\ell)} \big| 0 \right) &= \binom{\dv-1}{\f{0}^{(\ell)},  \f{1}^{(\ell)} , \dots, \f{\alpha^{q-2}}^{(\ell)}} \\
& \mkern-55mu \times  \left(1-\xi^{(\ell)}\right)^{\f{0}^{(\ell)}} \left(\xi^{(\ell)}/(q-1)\right)^{\dv-1-\f{0}^{(\ell)}}.
\end{align} 
Let us denote by $\I(\mathcal{P})$  the indicator function ($\I(\mathcal{P})$ takes value $1$ if the proposition $\mathcal{P}$ is true and $0$ otherwise). Let $\lmaxim{\ell}$ be the set of maximizers of $\aggv{\ell}$, i.e.,
\[
\lmaxim{\ell} = \left\{ \genfq \in \Fq \big| \aggc{\genfq}{\ell} = \max_{\a\in \Fq} \aggc{\a}{\ell}\right\} .
\] 
We may write 
\begin{equation} \label{eq:p_a2}
\vcm{0}{\ell} =   \sum_{\b \in \outputalphabet } \vcm{\b}{0}  \sum_{\fv^{(\ell-1)} }  P_{\Fv^{(\ell-1)}  | \inputrv } \left( \fv^{(\ell-1)}  \big| 0 \right)
 \frac{\I \left( 0 \in \lmaxim{\ell}  \right)}{|\lmaxim{\ell}|}   .
\end{equation}
Due to symmetry, for any $\a \neq 0$ we have
\[
\vcm{\a}{\ell} = \frac{1-\vcm{0}{\ell}}{q-1}.
\]
Note that, already for  moderate values of $q$ and $\dv$, the evaluation of \eqref{eq:p_a2} might be too complex.
In the Appendix, we provide tight upper and lower bounds on $\vcm{0}{\ell}$, which can be evaluated efficiently.
%by making use of a representation of the multinomial distribution as the sum of independent Poisson distributions, conditioned to their sum being fixed \cite{levin1981representation}. 
%Note that by upper bounding $\vcm{0}{\ell}$ we obtain an upper bound to the threshold.
%The threshold of an ensemble is obtained by iteratively applying \eqref{eq_c_v} and \eqref{eq:p_a2} for various $\proberror$.

%\begin{remark} 
%As a byproduct density evolution provides $\xi^{(\ell)} =1-\cvm{0}{\ell}$ which can be used as an estimate for the error probabilities \eqref{eq:LL-vector} and \eqref{eq:app}  for finite length simulations.
%\end{remark} 

%% file: tex_files/results.tex
\section{Numerical Results}\label{sec:results}

 In Table \ref{table:3:5:qsc}  we give iterative decoding thresholds on the \ac{QSC} for the ensemble $\ensC{3}{5}{q}$ for various $q$.
As a comparison, iterative decoding thresholds from \cite{kurkoski2007density} are reported for the simplest setup with list size $c=1$. Despite the larger message alphabet size for the algorithm in \cite{kurkoski2007density} with list size $1$ (which includes an additional erasure symbol), \ac{SMP} yields better thresholds.\footnote{We remark that increasing the list size in \cite{kurkoski2007density} yields an improvement in thresholds at the price of a higher computational burden.} This is owing to the proper choice of the message weights, as a result of \ac{DE} analysis from \eqref{eq:extrinsic_reliability}. %The results reported in \cite{kurkoski2007density}  are obtained by assuming that the channel observation and the messages from the extrinsic channels have the same reliability
The table also reports the Shannon limit $\proberror_{\mathrm{Sh}}$ and the belief propagation (BP) threshold $\proberror_{\text{BP}}$ obtained through Monte Carlo simulations \cite{Mac_Kay_LDPC_GFq_1998}. 
We  remark that as $q$ grows, the iterative decoding thresholds  $\proberror^{\star}$, the BP thresholds $\proberror_{\text{BP}}$ and $\proberror_{\mathrm{Sh}}$ increase.

\ctable[
caption = Thresholds for $\ensC{3}{5}{q}$ for different $q$.,
label   = table:3:5:qsc,
pos     = t,
doinside=\footnotesize
]
{rcccl}{}
{ 
\toprule
  $q$ &   $\proberror^{\star}$, \ac{SMP} & $\proberror^{\star}$ \cite{kurkoski2007density}, list size $1$ & $\proberror_{\text{BP}}$ & $\proberror_{\text{Sh}}$ \\ \midrule
  2 &   0.061   & 0.061& 0.113  &0.146 \\
  4 &   0.123   & 0.092& 0.196 &0.248 \\
  8 &   0.134   & 0.093& 0.254 &0.319 \\
  16 &   0.138  & 0.094& 0.296 &0.371 \\
  32 &   0.140  & --  & 0.328 &0.409 \\
  64 &   0.141  & --  &0.352&0.437 \\
  128 &  0.142  & --  & 0.371&0.459 \\
  256 &  0.142  & --   & 0.385&0.476 \\
  512 &  0.142  & --   &0.398&0.489 \\ \bottomrule
}

\ctable[
caption =  Thresholds for various rate-$1/2$ ensembles and different $q$.,
label   = table:3:6:qsc, 
pos     = t,
doinside=\footnotesize
]
{rccccl}{}
{	 
	\toprule
 $q$&   $\ensC{3}{6}{q}$ & $\ensC{4}{8}{q}$   & $\ensC{5}{10}{q}$   & $\ensC{6}{12}{q}$ &  $\proberror_{\text{Sh}}$ \\  \midrule
  2 &   0.040   & 0.052 & 0.042 & 0.040 &0.110 \\
  4 &   0.089   & 0.081 & 0.081 & 0.074 &0.189 \\
  8 &   0.104   & 0.106 & 0.101 & 0.101 &0.247\\
  16 &   0.108  & 0.137 & 0.116 & 0.112 &0.290\\
  32 &   0.109  & 0.164 & 0.136 & 0.121 &0.322\\
  64 &   0.110  & 0.176 & 0.162 & 0.135 &0.346\\
  128 &  0.111  & 0.182 & 0.177 & 0.156 &0.365\\
  256 &  0.111  & 0.185 & 0.185 & 0.170 &0.381\\
  512 &  0.111  & 0.186 & 0.188 & 0.178 &0.393\\ 
  \bottomrule
}

%\ctable[
%caption = Thresholds for the ensembles $\ensC{3}{6}{q}$, $\ensC{4}{8}{q}$, $\ensC{5}{10}{q}$ and $\ensC{6}{12}{q}$ for different $q$,
%label   = table:3:6:qsc,
%pos     = t
%]
%{rccccl}{}
%{	 
%	\toprule
% $q$&   $\ensC{3}{6}{q}$ & $\ensC{4}{8}{q}$   & $\ensC{5}{10}{q}$   & $\ensC{6}{12}{q}$ &  $\proberror_{\text{Sh}}$ \\  \midrule
%  2 &   0.040   & 0.052 & 0.042 & 0.040 &0.110 \\
%  4 &   0.089   & 0.081 & 0.081 & 0.074 &0.189 \\
%  8 &   0.104   & 0.106 & 0.101 & 0.101 &0.247\\
%  16 &   0.108  & 0.137 & 0.116 & 0.112 &0.290\\
%  32 &   0.109  & 0.164 & 0.136 & 0.121 &0.322\\
%  64 &   0.110  & 0.176 & 0.162 & 0.135 &0.346\\
%  128 &  0.111  & 0.182 & 0.177 & 0.156 &0.365\\
%  256 &  0.111  & 0.185 & 0.185 & 0.170 &0.381\\
%  512 &  0.111  & 0.186 & 0.188 & 0.178 &0.393\\ 
%}

Table~\ref{table:3:6:qsc} shows thresholds for $\ensC{3}{6}{q}$, $\ensC{4}{8}{q}$, $\ensC{5}{10}{q}$, and $\ensC{6}{12}{q}$ ensembles over the \ac{QSC} for different values of $q$. Note that the bounding techniques in the Appendix allow computing thresholds for large $q$, far beyond the values presented in the table. The ultra-sparse ensemble $\ensC{2}{4}{q}$ is not listed here, owing to a zero decoding threshold on the \ac{QSC}. %\footnote{The extrinsic output of a \ac{VN}  is determined by the channel message and one \ac{CN} message which, for $\proberror \in (0,1)$ will not overrule the channel vote.} 
For the binary case, the thresholds coincide with those achieved by the Gallager B algorithm. In fact, it is easy to recognize that \ac{SMP} with $q=2$ reduces, over the \ac{BSC}, to the Gallager B algorithm.
Interestingly, there seems to be no single regular \ac{LDPC} code ensemble with rate-$1/2$ that outperforms all others in terms of decoding threshold for all $q$. %For instance, for $q=4$, $\ensC{3}{6}{q}$ yields the best threshold, while for $q=512$, it is $\ensC{5}{10}{q}$.

\begin{figure}[t]
        \centering
        \includegraphics[width=0.98\columnwidth]{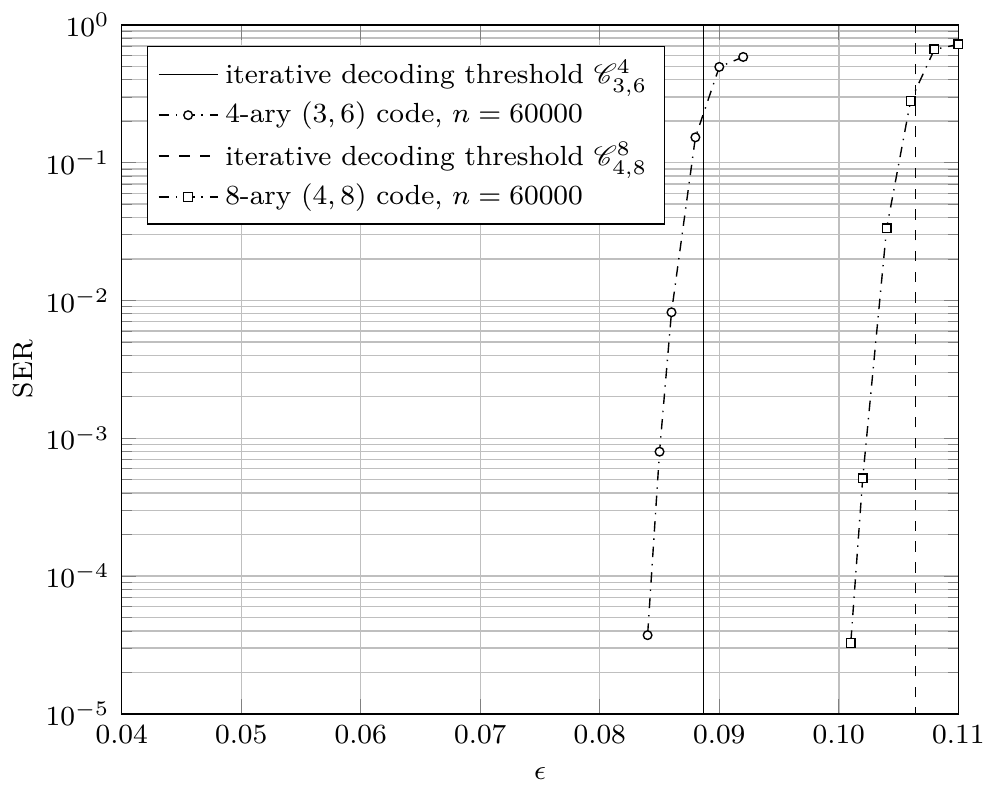}
        \caption{SER vs.\ channel error probability $\proberror$ for a $4$-ary $(3, 6)$ \ac{LDPC} code and a $8$-ary $(4, 8)$ \ac{LDPC} code with $n=60000$.}
        \label{fig:ser}
\end{figure}

Fig.~\ref{fig:ser} compares the iterative decoding threshold for the $\ensC{3}{6}{4}$  and $\ensC{4}{8}{8}$ \ac{LDPC} code ensembles with the \ac{SER} of a $4$-ary $(\dv=3,\dc=6)$  and $8$-ary $(\dv=4,\dc=8)$ \ac{LDPC} code, respectively, with $n=60000$.  The \ac{SER} results were obtained by Monte Carlo simulations and $200$ decoding iterations. As expected, the iterative decoding threshold predicts accurately the waterfall performance of the codes.

%% file: tex_files/conclusions.tex
\section{Conclusions}\label{sec:conclusions}

We presented symbol message passing, a low-complexity decoding algorithm for $q$-ary \ac{LDPC} codes. A \ac{DE} analysis is presented for regular ensembles over the \ac{QSC}. It yields iterative decoding thresholds and message weights which result in performance advantages with respect to a competing scheme of similar complexity. %focus on regular \ac{LDPC} code ensembles, although the extension to irregular ensembles is straightforward. 
 We also derived tight upper and lower bounds on the \ac{VN} message error probabilities, which allow efficient and accurate computation of the thresholds.

%% file: tex_files/appendix.tex
\section*{Efficient Evaluation of Density Evolution}\label{sec:appendix_efficient_DE}
We derive tight upper and lower bounds on \eqref{eq:p_a2}, which can be efficiently evaluated. For the sake of simplicity, whenever possible we drop the iteration count in the following. Let $ \nel{j} (\Fv) $ denote the number of elements of $\Fv$ equal to $j$, i.e., 
\begin{equation}
\nel{j} (\Fv) = \left|\left\{  \F{a}, a\in \Fq \big|  \F{a} = j\right\}\right| .
\end{equation}
Let us define $\accvote{b}$ as
\[
\accvote{b} = \f{b} + \frac{\D(\proberror)}{\D(\xi)}
\]
where we consider channels with non-zero capacity, i.e., $\D(\cdot)>0$.
Let \ac{VN} $\vn$ receive a channel message $y=0$, and $\f{0}$ messages with value $0$ from its neighbors. Whenever
\[
\max_{i \in \Fq \setminus 0} (\F{i}) < \az  
\]
the outgoing \ac{VN}-to-\ac{CN} message will be $0$. Further, whenever
\[
\max_{i \in \Fq \setminus 0} (\F{i}) = \az  % \qquad \mathrm{, and}  \qquad  \nel{\az}(\Fv) = \nmaxr
\]
the outgoing \ac{VN}-to-\ac{CN} message will take value $0$ with probability 1$/\nel{\az}(\Fv)$. %\fran{1$/(\nmaxr+1)$}
%Assume now that  $\vn$ has a channel observation $y=1$, and that it receives $\f{1}$ messages with value $1$, and let 
%\[
%\ao = \f{1} + \frac{\D(\proberror)}{\D(\xi)}.
%\]
%Whenever
%\[
%\max_{i=0, \alpha,\hdots, \alpha^{q-2}} (\F{i}) < \fmax
%\]
%the outgoing \ac{VN} to \ac{CN} message will be (always) $1$. 
%
%We have that whenver \
Similar considerations can be made when $y \neq 0$. Thus, for $q > 2$, we may recast \eqref{eq:p_a2}. This yields \eqref{eq:p0_expanded_u}, where
%\GLC{To be checked, after the fixes applied to (8):}
\begin{align}
\nmaxrmax &= \min \left( \Big\lfloor \frac{\dv-1 -\f{0}}{\az}  \Big\rfloor, q -1\right) \label{eq:max1} \\
\nmaxrmax' &= \min \left( \Big\lfloor \frac{\dv-1 -\f{1}}{\f{0}}  \Big\rfloor, q -1\right) \label{eq:max2} \\
\nmaxrmax'' &= \min \left( \Big\lfloor \frac{\dv-1 }{\ao} \Big\rfloor, q-1 \right)  \label{eq:max3}  .
\end{align}
%\[
%\ao = \f{1} + \frac{\D(\proberror)}{\D(\xi)}.
%\]

\newcounter{MYtempeqncnt}
\begin{figure*}[!t]
	% ensure that we have normalsize text
	\normalsize
	% Store the current equation number.
	\setcounter{MYtempeqncnt}{7}
	% Set the equation number to one less than the one
	% desired for the first equation here.
	% The value here will have to changed if equations
	% are added or removed prior to the place these
	% equations are referenced in the main text.
	\setcounter{equation}{\value{equation}}
	\begin{align} 
	\vcm{0}{\ell}  & = \vcm{0}{0}\sum_{\f{0}=0}^{\dv-1} P_{\F{0}| \inputrv} (\f{0}| 0)
	\Bigg(\Pr \left\{  \max_{i \in \Fq \setminus 0} (\F{i}) < \az \big| \inputrv =0, \F{0} = \f{0} \right\} \hfill \\
		 	&+  \Pr \left\{  \max_{i \in \Fq \setminus 0} (\F{i}) = \az  \big| \inputrv =0, \F{0} = \f{0} \right\} 
		 	\underbrace{\sum_{\nmaxr=1}^{\nmaxrmax} \frac{1}{\nmaxr+1} \Pr\left\{ \nel{\az} (\Fv) = \nmaxr \big| X=0, \F{0} = \f{0}, \max_{i \in \Fq \setminus 0} (\F{i})= \az  \right\}}_{\text{(a)}     } 	
	\Bigg) \\	
	&  +  (q-1)\, \vcm{1}{0}   \sum_{\f{1}=0}^{\dv-1} P_{\F{1}|\inputrv} (\f{1}|0)
	 \Bigg[ \sum_{\f{0}> \ao }^{\dv-1-\f{1}}   P_{\F{0}| \inputrv, \F{1}} (\f{0}  | 0, \f{1} )
	 \Bigg( \Pr \left\{  \max_{i \in \Fq \setminus \{0,1\}} (\F{i}) < \f{0} \big| \inputrv=0, \F{0} = \f{0}, \F{1} = \f{1} \right\} \\[-0.25em]
	&+ \Pr \left\{  \max_{i \in \Fq \setminus \{0,1\}} (\F{i}) = \f{0} \bigg| \inputrv=0, \F{0} = \f{0}, \F{1} = \f{1} \right\}   \\
	  & \times \underbrace{\sum_{\nmaxr=2}^{\nmaxrmax'} \frac{1}{\nmaxr }
	\Pr\left\{\nel{\f{0}}(\Fv) = \nmaxr  \bigg| \inputrv=0,  \F{0} = \f{0}, \F{1} = \f{1}, \max_{i \in \Fq \setminus \{0,1\}} (\F{i}) = \f{0}  \right\}}_{\text{(b)}}  \Bigg)	\\[-0.25em]
	&+ P_{\F{0}| \inputrv, \F{1}} ( \ao  | 0, \f{1} ) 
	\Bigg( \frac{1}{2} \Pr \left\{  \max_{i \in \Fq \setminus \{0,1\}} (\F{i}) < \ao \bigg|\inputrv=0, \F{0} = \ao, \F{1} = \f{1} \right\} \\[-0.25em]
	&
	+ \Pr \left\{ \max_{i \in \Fq \setminus \{0,1\}} (\F{i}) = \ao \bigg| \inputrv=0, \F{0} = \ao, \F{1} = \f{1} \right\} %\\
	%&  \phantom{=} &&\times 
	   \\
	& \times \underbrace{\sum_{\nmaxr=2}^{\nmaxrmax''} \frac{1}{\nmaxr+1}
	\Pr\left\{ \nel{\ao} (\Fv)= \nmaxr  \bigg|\inputrv=0, \F{0} = \ao, \F{1} = \f{1},  \max_{i \in \Fq \setminus \{0,1\}} (\F{i}) = \ao \right\}}_{\text{(c)}}  \Bigg) \Bigg]	\label{eq:p0_expanded_u} 
	\end{align}

	% Restore the current equation number.
	\setcounter{equation}{8}
	% IEEE uses as a separator
	\hrulefill
	% The spacer can be tweaked to stop underfull vboxes.
	\vspace*{4pt}
\end{figure*}

An upper bound on $\vcm{0}{\ell}$ is obtained as follows. Whenever the aggregated \acl{llv}  $\aggv{\ell}$ has $\nmaxr>1$ maxima, one of them being at $0$, we assume that $\aggv{\ell}$ has the minimum possible number of maxima. We thus replace the terms (a), (b), (c) in \eqref{eq:p0_expanded_u} by $1/2$, $1/2$, and $1/3$, respectively. Similarly, a lower bound can be obtained by replacing the terms (a), (b), (c) in \eqref{eq:p0_expanded_u} by $1/(\nmaxrmax+1)$,  $1/ \nmaxrmax'$, and  $1/( \nmaxrmax''+1)$, respectively. %from \eqref{eq:max1}, \eqref{eq:max2}, \eqref{eq:max3}, respectively. 
For the lower bound we thus overestimate the number of maxima. Both upper and lower bounds can be efficiently evaluated using a result in \cite{levin1981representation}. Both bounds are tight for the ensembles in Tables \ref{table:3:5:qsc} and \ref{table:3:6:qsc}. In fact, they coincide in the first $6$ decimal digits.

%% file: main.bbl
\begin{thebibliography}{10}
\providecommand{\url}[1]{#1}
\csname url@samestyle\endcsname
\providecommand{\newblock}{\relax}
\providecommand{\bibinfo}[2]{#2}
\providecommand{\BIBentrySTDinterwordspacing}{\spaceskip=0pt\relax}
\providecommand{\BIBentryALTinterwordstretchfactor}{4}
\providecommand{\BIBentryALTinterwordspacing}{\spaceskip=\fontdimen2\font plus
\BIBentryALTinterwordstretchfactor\fontdimen3\font minus
  \fontdimen4\font\relax}
\providecommand{\BIBforeignlanguage}[2]{{%
\expandafter\ifx\csname l@#1\endcsname\relax
\typeout{** WARNING: IEEEtran.bst: No hyphenation pattern has been}%
\typeout{** loaded for the language `#1'. Using the pattern for}%
\typeout{** the default language instead.}%
\else
\language=\csname l@#1\endcsname
\fi
#2}}
\providecommand{\BIBdecl}{\relax}
\BIBdecl

\bibitem{Gallager63}
R.~G. Gallager, ``Low-density parity-check codes,'' Ph.D. dissertation, Dep.
  Electrical Eng., M.I.T, Cambridge, MA, Jul. 1963.

\bibitem{richardson2001capacity}
T.~Richardson and R.~Urbanke, ``The capacity of low-density parity-check codes
  under message-passing decoding,'' \emph{{IEEE} Trans. Inf. Theory}, vol.~47,
  no.~2, pp. 599--618, Feb. 2001.

\bibitem{Lechner:BMP}
G.~{Lechner}, T.~{Pedersen}, and G.~{Kramer}, ``Analysis and design of binary
  message passing decoders,'' \emph{{IEEE} Trans. Commun.}, vol.~60, no.~3, pp.
  601--607, Mar. 2012.

\bibitem{yacoub2018protograph}
E.~B. Yacoub, F.~Steiner, B.~Matuz, and G.~Liva, ``Protograph-based {LDPC} code
  design for ternary message passing decoding,'' in \emph{Proc. ITG Int. Conf.
  Syst., Commun. and Coding}, Rostock, Germany, Feb. 2019.

\bibitem{planjery2013finite}
S.~K. Planjery, D.~Declercq, L.~Danjean, and B.~Vasic, ``Finite alphabet
  iterative decoders -- part i: Decoding beyond belief propagation on the
  binary symmetric channel,'' \emph{{IEEE} Trans. Commun.}, vol.~61, no.~10,
  pp. 4033--4045, Oct. 2013.

\bibitem{Mac_Kay_LDPC_GFq_1998}
M.~Davey and D.~MacKay, ``Low density parity check codes over {GF}$(q)$,''
  \emph{{IEEE} Commun. Lett.}, vol.~2, no.~6, pp. 70--71, Jun. 1998.

\bibitem{Declerq:FFT_dec_2003}
L.~Barnault and D.~Declercq, ``Fast decoding algorithm for {LDPC} over
  {GF}($2^q$),'' in \emph{Proc. IEEE Inf. Theory Workshop (ITW)}, Cergy,
  France, Mar. 2003, pp. 70--73.

\bibitem{declercq2007decoding}
D.~Declercq and M.~Fossorier, ``Decoding algorithms for nonbinary {LDPC} codes
  over {GF ($q$)},'' \emph{{IEEE} Trans. Commun.}, vol.~55, no.~4, pp.
  633--643, Apr. 2007.

\bibitem{metzner1996majority}
J.~J. Metzner, ``Majority-logic-like decoding of vector symbols,'' \emph{{IEEE}
  Trans. Commun.}, vol.~44, no.~10, pp. 1227--1230, Oct. 1996.

\bibitem{luby2005verification}
M.~G. Luby and M.~Mitzenmacher, ``Verification-based decoding for packet-based
  low-density parity-check codes,'' \emph{{IEEE} Trans. Inf. Theory}, vol.~51,
  no.~1, pp. 120--127, Jan. 2005.

\bibitem{MLP+13}
B.~{Matuz}, G.~{Liva}, E.~{Paolini}, and M.~{Chiani}, ``Verification-based
  decoding with map erasure recovery,'' in \emph{Proc. ITG Int. Conf. Syst.
  Commun. and Coding}, M\"unchen, Germany, Jan 2013.

\bibitem{shokrollahi2004low}
A.~Shokrollahi and W.~Wang, ``Low-density parity-check codes with rates very
  close to the capacity of the q-ary symmetric channel for large q,'' in
  \emph{Proc. IEEE Int. Symp. on Inf. Theory}, Jun. 2004, p. 273.

\bibitem{zhang2011analysis}
F.~Zhang and H.~D. Pfister, ``Analysis of verification-based decoding on the
  $q$-ary symmetric channel for large $q$,'' \emph{{IEEE} Trans. Inf. Theory},
  vol.~57, no.~10, pp. 6754--6770, 2011.

\bibitem{kurkoski2007density}
B.~M. Kurkoski, K.~Yamaguchi, and K.~Kobayashi, ``Density evolution for
  {GF($q$) LDPC} codes via simplified message-passing sets,'' in \emph{Proc,
  IEEE Inf. Theory and Appl. Workshop}, 2007, pp. 237--244.

\bibitem{stark2018decoding}
M.~Stark, G.~Bauch, J.~Lewandowsky, and S.~Saha, ``Decoding of non-binary
  {LDPC} codes using the information bottleneck method,'' in \emph{Proc. IEEE
  Int. Conf. on Commun.}, 2019.

\bibitem{mdpc}
R.~{Misoczki}, J.~{Tillich}, N.~{Sendrier}, and P.~S. L.~M. {Barreto},
  ``{MDPC-McEliece}: New {McEliece} variants from moderate density parity-check
  codes,'' in \emph{Proc of IEEE Intern. Symp. on Inf. Theory}, Jul. 2013.

\bibitem{ashikhmin2004extrinsic}
A.~Ashikhmin, G.~Kramer, and S.~ten Brink, ``Extrinsic information transfer
  functions: model and erasure channel properties,'' \emph{{IEEE} Trans. Inf.
  Theory}, vol.~50, no.~11, pp. 2657--2673, Nov. 2004.

\bibitem{lazaro2018bounds}
F.~L{\'a}zaro, G.~Liva, E.~Paolini, and G.~Bauch, ``Bounds on the error
  probability of {Raptor} codes under maximum likelihood decoding,''
  \emph{arXiv preprint arXiv:1809.01515}, 2018.

\bibitem{levin1981representation}
B.~Levin, ``A representation for multinomial cumulative distribution
  functions,'' \emph{The Annals of Statistics}, vol.~9, no.~5, pp. 1123--1126,
  1981.

\end{thebibliography}
